\documentclass[11pt,twoside]{article}

\usepackage{asp2006}
\usepackage{epsf}
\usepackage{lscape}

\def\sun{\hbox{$\odot$}}
\def\R23{\mbox{$\rm R_{23}$}}

\def\Hb{\mbox{${\rm H}{\beta}$}}

\def\OIIIa{\mbox{${\rm [O\,III]\,}{\lambda\,5007}$}}

\def\OII{\mbox{${\rm [O\,II]\,}{\lambda\,3727}$}}

\begin{document}
\title{The evolution of the star formation of zCOSMOS and SDSS galaxies at $z<0.7$
as a function of mass and structural parameters}   %%% Fill in title
\author{Christian Maier}   %%% Fill in author names
\affil{Institute of Astronomy, ETH Z\"urich, CH-8093, Z\"urich, Switzerland}    %%% Fill in author affiliations
\author{and the zCOSMOS Collaboration}   %%% Fill in author names

\begin{abstract} %%% Abstract to run on from here.
We present in these proceedings some preliminary results we have obtained studying
the evolution of the specific star formation rate as a function of
surface mass density and Sersic indices at $z<0.7$.
These results are based on the consistent  comparison of the properties of $\sim$ 650 massive
zCOSMOS galaxies in a mass-complete sample at $0.5<z<0.7$ with a
mass-complete sample of $\sim 21500$ SDSS
local galaxies.
\end{abstract}

%%% MAIN BODY OF TEXT GOES HERE. CONSULT "INSTRUCTIONS FOR AUTHORS USING
%%% LATEX2E MARKUP", SECTIONS 2.3-2.6 FOR HELP WITH EQUATIONS, FIGURES,
%%% AND TABLES.

\section{Introduction}   %%% Top level section head (remove "%" symbol)

One of the key unanswered questions is what physical parameter(s) drive
changes in the star formation rate in individual galaxies.
Given the strong correlation (except at the highest masses) between star formation rate (SFR) and
stellar mass \citep[see, e.g., Fig.17 in][]{brinchmann04}, one can more
easily study the relationship between star formation activity and the
physical parameters of galaxies by normalizing the SFR by stellar mass.
The star formation rate per unit stellar mass, the specific star
formation rate (SSFR), is  an  indicator of 
galaxy star formation histories, since 1/SSFR defines a characterisic
time-scale of star formation.

 In the local
Universe, there are several studies of the role of stellar mass, $M_{*}$, and
stellar mass density, $\Sigma_{M}$, in
regulating the star formation activity.
Using the SDSS sample \citet{brinchmann04}  found that a low SSFR peak becomes more
prominent at high $\Sigma_{M}$ than at high $M_{*}$, and concluded
that the surface density of stars
is more important than  stellar mass in determining when star formation
is turned off.

We present here  some preliminary results on the role of the stellar mass surface
density in regulating the star formation activity at $z<0.7$ for galaxies of different morphologies. 
This study is based on a $0.5<z<0.7$ mass-complete sample
of zCOSMOS galaxies. The redshift range $0.7<z<0.9$ and more discussion
of our results will be addressed in \citet{maier09}.

%########################################################

\section{Selection of the $0.5<z<0.7$ mass-complete zCOSMOS sample  and the consistent
  determination of physical properties for zCOSMOS and SDSS}

The Cosmic Evolution Survey (COSMOS) is designed to probe the
correlated evolution of galaxies, star formation, active galactic
nuclei (AGNs), and dark matter with large-scale structure up to high
redshifts. COSMOS is the largest HST survey ever undertaken, imaging an
equatorial, $\sim 2 \rm{deg}^{2}$ field, with single-orbit I-band exposures.
Ancillary  COSMOS data  from  Spitzer, XMM-Newton,
VLA, and a large number of other ground-based telescopes are
available. 

 zCOSMOS \citep{lilly07} is a large redshift survey that is being undertaken in the
COSMOS field using 600 hours of observation with the VIMOS spectrograph
on the VLT.
The zCOSMOS survey consists of two parts: a) zCOSMOS-bright, a magnitude-limited I-band I$_{AB}<22.5$ sample of
about 20\,000 galaxies with $0.1<z<1.2$ covering the whole 1.7\,deg$^{2}$
COSMOS ACS field;
and b) zCOSMOS-deep, a survey of approximately 10\,000 galaxies selected
through colour-selection criteria to have $1.4<z<3.5$, within the
central 1 deg$^{2}$ of the COSMOS field.

This study is based on the zCOSMOS-bright sample, and the $0.5<z<0.7$ galaxies are
selected from 83 VIMOS masks.
Line fluxes from the spectra were measured using the automatic software
Platefit \citep{lamar08}.
For galaxies where Platefit detects all three emission
lines \OIIIa, \Hb, and \OII,  
we used the \OIIIa/\Hb\, vs. \OII/\Hb\, diagram  to disentangle star-formation dominated galaxies from
objects obviously
containing an active nucleus (narrow-line AGNs) using   the
empirical threshold   derived using the 2dFGRS by \citet{lamar04}.
Additionally,  the X-ray identified AGNs were excluded.

Physical sizes  of the zCOSMOS galaxies are computed from the
half-light radii derived from GIM2D Sersic fits \citep{sargent07} on the ACS images.
Stellar masses were derived  using the relation between 
rest-frame U-B and B-V colors and mass-to-light ratio (M/L), using the
equation (1) from \citet{lin07}, which corrects M/L  for 
redshift evolution, as well as accounting for evolution in color.
The \OII\, line luminosities corrected for aperture effects are transformed into SFRs using a
correction factor based on the galaxy's B-band absolute magnitude, as
given by a linear interpolation of the values in Table 2 of \citet{moust06},
and shown in their Fig.\,19.

We used \citet{bruzcharl03} models to assess the mass completeness of
the zCOSMOS sample. 
It turned out that a selection of $\rm{logM}_{*}>10.4$ for $0.5<z<0.7$ zCOSMOS galaxies is
required to obtain a mass-complete sample which does not miss low mass
early-type galaxies.
This way we remained with  $\sim 650$ zCOSMOS galaxies at $0.5<z<0.7$.

We would like to point out that masses (from colors), sizes (from
g-band images), Sersic indices (from g-band images), and SFRs (from the
\OII\, line flux) were computed
in a consistent way for the zCOSMOS and SDSS mass-complete samples.
More details on the consistent derivation of physical parameters for
the zCOSMOS and SDSS sample can be found in \citet{maier09}.

%#######################################################################################

\section{Specific star formation rate (SSFR) versus stellar mass
  surface density ($\Sigma_{M}$) at $z<0.7$}

Fig.\,1 shows the SSFR vs. stellar mass surface density 
for the mass-complete sample ($\rm{logM}_{*}>10.4$) of  zCOSMOS
galaxies at $0.5<z<0.7$, and for $\rm{logM}_{*}>10.4$ SDSS
galaxies at $0.04<z<0.08$, as a function of Sersic
index $n$.
We show in Fig.\,1  only galaxies with an axis ratio $b/a>0.55$, because of the
lower inclination of these objects, which
minimizes the effects of dust extinction on galaxy colors, used to
derive stellar masses, and
on the \OII\, line flux, used to derive star formation rates.

The individual measurements for zCOSMOS $0.5<z<0.7$ galaxies with
$n<1.5$ (disk galaxies) and $n>2.5$ (early-type galaxies) are shown
as cyan dots in panel a) and b), while panel c) compares the median
values of zCOSMOS and SDSS  mass-complete samples of
$b/a>0.55$ galaxies (open symbols) with the median values shown in
panel a) and b) for low and
high Sersic indices (filled squares for zCOSMOS, filled triangles for SDSS).
Diagonal dashed lines correspond to SFR surface densities
$\Sigma_{SFR} = 0.01 M_{\sun}/\rm{yr/kpc}^{2}$, and $0.1 M_{\sun}/\rm{yr/kpc}^{2}$, respectively.
%
%##########################################################################

\begin{figure}
\begin{center}
\hspace{-1.4cm}
\resizebox{13cm}{!}{\includegraphics{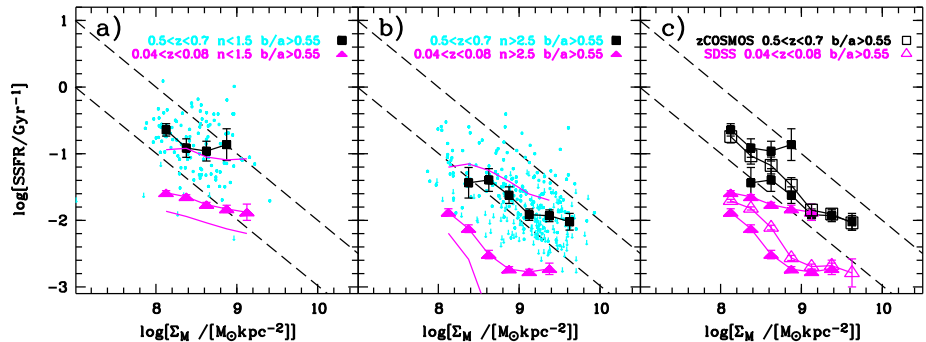}}\\%
\caption{ \footnotesize 
Specific star formation rate vs. stellar mass surface density for the
zCOSMOS mass complete sample at
$0.5<z<0.7$ (individual measurements shown as cyan dots, and median values shown as black squares) compared
to the median SSFR values in different $\Sigma_{M}$ bins of $0.04<z<0.08$ SDSS
galaxies (magenta triangles).
Moreover,  the 25th and 95th percentiles of
SSFR in the respective SDSS  $\Sigma_{M}$ bin are shown as solid magenta lines.
Diagonal dashed lines correspond to SFR surface densities
$\Sigma_{SFR} = 0.01 M_{\sun}/\rm{yr/kpc}^{2}$, and $0.1 M_{\sun}/\rm{yr/kpc}^{2}$, respectively.
For $n<1.5$ galaxies (panel a) the SSFR stays roughly constant
(or decreases slightly) with $\Sigma_{M}$, with the median SSFR at a given
$\Sigma_{M}$ being higher in zCOSMOS than in SDSS.
For $n>2.5$ galaxies (panel b) the SSFR declines with  $\Sigma_{M}$ for both
zCOSMOS and SDSS galaxies, but there is a shift to higher  $\Sigma_{M}$
for zCOSMOS galaxies, additional to the fact that the SSFR of zCOSMOS
galaxies is higher than for SDSS.
The median values of all  $b/a>0.55$ galaxies in the  mass-complete samples of SDSS and zCOSMOS galaxies are
shown as open symbols in panel c.
}
\label{Figure1}
\end{center}
\end{figure}

%##########################################################################

%
The 25th and 95th percentiles of the
SSFR in the respective SDSS  $\Sigma_{M}$ bin are  shown as solid
magenta lines in panel a) and b).
The decrease in the  25th
percentile of SSFR we see for SDSS $n>2.5$ galaxies  suggests that there is
an increasing supression of star
formation as the bulge of the galaxy becomes more dominant. This is
unlike the situation for
$n<1.5$ SDSS galaxies, where the  25th percentile of the SSFR distribution
decreases only slightly, which is due to the fact that most $n<1.5$
galaxies have elevated levels of star formation.

For both  zCOSMOS and SDSS $n<1.5$ galaxies (panel a) the SSFR stays roughly constant
(or decreases slightly) with $\Sigma_{M}$, with the median SSFR at a given
$\Sigma_{M}$ being higher in zCOSMOS than in SDSS.
A declining SFR in zCOSMOS $n<1.5$ (disk) galaxies would explain their
evolution in order to reach the location of SDSS $n<1.5$ galaxies today.
For $n>2.5$ (early-type) galaxies (panel b) the SSFR declines with  $\Sigma_{M}$ for both
zCOSMOS and SDSS galaxies, but there is a shift to higher  $\Sigma_{M}$
for zCOSMOS galaxies, additional to the fact that the SSFR of zCOSMOS
galaxies is higher than for SDSS.
Panel c) shows how the different behaviour of median values for low and
high Sersic index galaxies can explain the general trend seen in the
population (open symbols).
A more in-depth discussion of these interesting results and additional analysis of
the $0.7<z<0.9$ zCOSMOS mass-complete sample is presented in \citet{maier09}.

%##########################################################################

%%%\acknowledgements %%% Text of acknowledgements runs on after this command.

%%% THE BIBLIOGRAPHY
%%%
%%% CONSULT SECTION 3 OF "INSTRUCTIONS FOR AUTHORS" FOR HOW TO USE NATBIB.
%%% AUTHORS ARE ENCOURAGED TO USE EITHER THE "THEBIBLIOGRAPY" ENVIRONMENT
%%% BY UNCOMMENTING (DELETING THE "%" SYMBOL) THE COMMANDS BELOW, OR BY
%%% USING THE BIBTEX ENVIRONMENT. TO FIND OUT WHICH IS APPLICABLE TO YOUR
%%% CONTRIBUTION, CONSULT THE VOLUME EDITORS FOR YOUR PROCEEDINGS.
%%%


\begin{thebibliography}{}


\bibitem[Brinchmann et al.(2004)]{brinchmann04} Brinchmann, J., Charlot, S., White, S.D.M. et al.  2004, MNRAS, 351, 1151 

\bibitem[Bruzual \& Charlot(2003)]{bruzcharl03} Bruzual, G., \&
  Charlot, S., 2003, MNRAS, 344, 1000


\bibitem[Lamareille et al.(2004)]{lamar04} Lamareille, F., Mouhcine,
  M., Contini, T., Lewis, I. \& Maddox, S., 2004, MNRAS, 350, 396

\bibitem[Lamareille et al.(2008)]{lamar08} Lamareille, F. et al. 2008, in preparation

\bibitem[Lilly et al.(2007)]{lilly07} Lilly S.J. et al. 2007, ApJS, 172, 70

\bibitem[Lin et al.(2007)]{lin07} Lin L., Koo D.C., Weiner B.J., et al. 2007, ApJ, 660, 51

\bibitem[Maier et al.(2009)]{maier09} Maier et al. 2009, ApJ, 694, 1099

\bibitem[Moustakas et al.(2006)]{moust06} Moustakas, J., Kennicutt,
  Jr., R. C., \& Tremonti, C. A., 2006, ApJ, 642, 775

\bibitem[Sargent et al.(2007)]{sargent07} Sargent, M. et al., 2007,
  ApJS, 172, 434

\end{thebibliography}
\end{document}